 \documentclass[final,5p,times,twocolumn]{elsarticle}

 \biboptions{numbers,sort&compress}

\usepackage{amssymb}
\usepackage{lipsum}
\usepackage{graphicx}% Include figure files
\usepackage{dcolumn}% Align table columns on decimal point
\usepackage{bm}% bold math
\usepackage{braket}
\usepackage{subcaption}
\usepackage{amsmath}
\usepackage{mathrsfs}
\usepackage{subcaption}
\usepackage{ragged2e}
\usepackage{gensymb}
\usepackage{comment}
\usepackage{hyperref}
\usepackage{cleveref}
\usepackage{soul}

\usepackage[T1]{fontenc}
\usepackage{tikz}
\usepackage{quantikz}

\usetikzlibrary{calc}

\usetikzlibrary{decorations.pathmorphing}

\journal{Physics Letters B}

\begin{document}

\begin{frontmatter}

%% Title, authors and addresses

%% use the tnoteref command within \title for footnotes;
%% use the tnotetext command for theassociated footnote;
%% use the fnref command within \author or \affiliation for footnotes;
%% use the fntext command for theassociated footnote;
%% use the corref command within \author for corresponding author footnotes;
%% use the cortext command for theassociated footnote;
%% use the ead command for the email address,
%% and the form \ead[url] for the home page:
%% \title{Title\tnoteref{label1}}
%% \tnotetext[label1]{}
%% \author{Name\corref{cor1}\fnref{label2}}
%% \ead[url]{home page}
%% \fntext[label2]{}
%% \cortext[cor1]{}
%% \affiliation{organization={},
%%            addressline={}, 
%%            city={},
%%            postcode={}, 
%%            state={},
%%            country={}}
%% \fntext[label3]{}

\title{Quantum simulation of neutrino oscillations with bosonic encoding}

%% use optional labels to link authors explicitly to addresses:
%% \author[label1,label2]{}
%% \affiliation[label1]{organization={},
%%             addressline={},
%%             city={},
%%             postcode={},
%%             state={},
%%             country={}}
%%
%% \affiliation[label2]{organization={},
%%             addressline={},
%%             city={},
%%             postcode={},
%%             state={},
%%             country={}}

\author{Sandeep Joshi}
\ead{sjoshi@barc.gov.in}
\affiliation{organization={TNP\&QC Section, Nuclear Physics Division, Bhabha Atomic Research Centre},%Department and Organization
            city={Mumbai},
            postcode={400085}, 
            country={India}}

\begin{abstract}
%% Text of abstract
Superconducting qubits offer a versatile platform for quantum simulation. In this architecture, quantum information can be encoded in the bosonic modes of a microwave cavity, offering an alternative to conventional qubit-based encoding schemes. These cavity bosonic modes can be manipulated using a single ancillary qubit. In this work, we investigate the quantum simulation of two- and three-flavor neutrino oscillations using Fock-basis encoding of a cavity mode. We design pulse sequences for implementing the required unitary operations through selective number-dependent arbitrary phase (SNAP) and displacement gates. Pulse-level control is employed to realize high-fidelity gate operations on the encoded cavity mode. The resulting neutrino oscillation probabilities obtained from quantum simulation exhibit close agreement with the corresponding theoretical predictions, demonstrating the feasibility of cavity-based bosonic encoding schemes for quantum simulation.
\end{abstract}

%%Graphical abstract
%\begin{graphicalabstract}
%\includegraphics{grabs}
%\end{graphicalabstract}

%%Research highlights
%\begin{highlights}
%\item Research highlight 1
%\item Research highlight 2
%\end{highlights}

\begin{keyword}
%% keywords here, in the form: keyword \sep keyword, up to a maximum of 6 keywords
Quantum simulation \sep neutrino oscillations \sep bosonic encoding  \sep superconducing qubits

%% PACS codes here, in the form: \PACS code \sep code

%% MSC codes here, in the form: \MSC code \sep code
%% or \MSC[2008] code \sep code (2000 is the default)

\end{keyword}

\end{frontmatter}

%\tableofcontents

%% \linenumbers

%% main text

\section{Introduction}
\label{introduction}
Quantum simulation of nuclear and particle physics phenomena is emerging as an increasingly important research area. With the growing availability and enhanced performance of quantum hardware, simulation of otherwise intractable problems is becoming progressively more feasible \cite{Bauer:2023qgm, Bauer:2022hpo, DiMeglio:2023nsa}. In general, a complex quantum system can be simulated using a controllable quantum platform through either analog or digital approaches \cite{Lloyd:1996aai, Georgescu:2013oza}. In the case of digital quantum simulation, the unitary evolution of the target system is implemented via a sequence of quantum gates, thereby reproducing the dynamics of the original system.

Neutrino oscillation is an archetypal phenomenon that has served as an important testbed for quantum simulation. Owing to the distinctive features of neutrino oscillations, several studies have explored quantum simulation of different aspects of this phenomenon \cite{noh2012quantum,  Jha:2021itm, Joshi:2025jxe, Hall:2021rbv,Yeter-Aydeniz:2021olz, Balantekin:2023qvm, Amitrano:2022yyn,Tripathi:2025cok}. Quantum simulations of two- and three-flavor neutrino oscillations typically employ encoding schemes based on one and two qubits, respectively \cite{Arguelles:2019phs, Singh:2024vpu}. Alternatively, a qutrit can also be used to encode the three neutrino flavors \cite{Nguyen:2022snr, Spagnoli:2025etu}. In these approaches, the unitary time evolution is decomposed into sequences of single- and two-qubit gates, which are subsequently implemented on the encoded qubits. Measurement of the final qubit states then yields the neutrino oscillation probabilities.

Superconducting qubits have emerged as one of the most promising platforms for the implementation of quantum computing. This architecture consists of Josephson junction based qubits, such as transmons \cite{Koch:2007hay}, coupled to microwave cavities realized in either three-dimensional or planar two-dimensional structures. The engineered interaction between the qubits and the cavity microwave photons forms the basis of circuit quantum electrodynamics (QED) \cite{Blais:2020wjs}. In the prototypical circuit QED setup, the logical states are encoded in the transmon qubits. Logical gate operations are performed using microwave drive lines, with one drive line dedicated to each qubit, while the cavity is primarily employed for qubit state readout. However, transmons possess finite coherence times, and scaling to multiple qubits requires increasingly complex control hardware with several drive lines. Consequently, the implementation of large-scale logical gate operations becomes experimentally challenging. 

An alternative approach involves encoding quantum information in the bosonic modes of the microwave cavity. A single bosonic mode possesses an infinite-dimensional Hilbert space, from which a suitable $d$-dimensional logical subspace, referred to as a qudit, can be selected for information processing. Bosonic encodings offer several advantages, most notably the long coherence times of cavity modes. In addition, the control hardware requirements are comparatively simpler, as fewer control channels are needed for manipulating the entire qudit space. The use of cavity bosonic modes has attracted significant interest in applications ranging from quantum simulation \cite{Kurkcuoglu:2021dnw, Dutta:2024zep, Liu:2024mbr} to quantum error correction \cite{Joshi:2020sen, Cai:2020jvu}. Quantum information can be encoded in cavity bosonic modes using several different schemes such as cat states \cite{Vlastakis:2013aha}, binomial states \cite{Michael:2016eaq}. Among these approaches, one of the simplest is to encode information in the Fock state basis of the cavity \cite{Alam:2022crs}.

 Quantum control of cavity bosonic modes is a challenging task. In the circuit QED architecture, universal control over cavity states is achieved by employing transmons as ancillary nonlinear element. There are several techniques which have been proposed to realize universal control of the bosonic modes of the cavity \cite{Ma:2021rum}.  In this work, we employ the protocol introduced in \cite{Krastanov:2015rkt, Heeres:2015cnj}, which combines conditional ancilla rotations with cavity displacement operations to achieve arbitrary control over cavity Fock states.  The ancilla rotation is conditioned on the photon-number state of the cavity mode and is referred to as the SNAP gate. The displacement operation, in contrast, acts unconditionally on the cavity mode. By interleaving sequences of SNAP gates and displacement operations, universal control over the cavity Hilbert space can be realized.

This paper is organized as follows. In Section 2, we provide an overview of the bosonic encoding framework in circuit QED and the pulse-level implementation of quantum control using the SNAP-displacement protocol. In Section 3, we present pulse-level simulations of two-flavor neutrino oscillations. Section 4 is devoted to the simulation of three-flavor neutrino oscillations, where we implement an optimized displacement-SNAP gate sequence for realizing the required unitary dynamics. Finally, we summarize our results and conclude in Section 5.

\begin{figure*}[t]
\centering

\scalebox{1.0}{

\begin{tikzpicture}[remember picture]

% ======================
% QUANTIKZ CIRCUIT
% ======================
\node (circuit) {
\begin{quantikz}[column sep=0.9cm, row sep=0.3cm]
\lstick{$C$}
    & \gate[style={draw=purple, thick, rounded corners}]{D(\alpha_1)}
    & \ctrl{1}
    %& \gate[2, style={draw=olive, thick, rounded corners}]{S(\Vec{\theta}_1)}
    & \gate[style={draw=purple, thick, rounded corners}]{D(\alpha_2)}
    & \ctrl{1}
    %& \gate[2, style={draw=olive,  thick, rounded corners}]{S(\Vec{\theta}_2)}
    & \ \ldots \ 
    & \gate[style={draw=purple, thick, rounded corners}]{D(\alpha_k)}
    & \ctrl{1}
    %& \gate[2, style={draw=olive,  thick, rounded corners}]{S(\Vec{\theta}_k)}
    & \gate[style={draw=purple, thick, rounded corners}]{D(\alpha_{k+1})}
    & \qw \\
\lstick{$A$}
    & \qw 
    & \gate[style={draw=olive, thick, rounded corners}]{S(\Vec{\theta}_1)}
    & \qw 
    & \gate[style={draw=olive, thick, rounded corners}]{S(\Vec{\theta}_2)}
    & \ \ldots \   & \qw 
    & \gate[style={draw=olive, thick, rounded corners}]{S(\Vec{\theta}_k)}
    \qw & \qw & \qw
\end{quantikz}
};

% ======================
% PULSE SEQUENCES
% ======================

\begin{scope}[yshift=-3.8cm, xshift=-9cm, x=1.2cm, y=0.8cm]

% =====================================================
% TIME AXIS
% =====================================================

\draw[thick,->] (0.8,0) -- (14.5,0);

\node[above] at (0.4,-0.25) {Pulse};

\node[right] at (14.5,0) {\small time};

% =====================================================
% DISPLACEMENT GAUSSIAN PULSES
% =====================================================

\foreach \x in {2.15,5.25,9.9,13}
{
\filldraw[
fill=purple!35,
fill opacity=0.1,
purple,
thick,
smooth,
samples=100,
domain=-0.45:0.45,
variable=\u
]
plot
(
{\x+\u},
{2*exp(-22*\u*\u)}
);
}

% =====================================================
% SNAP FLAT-TOP GAUSSIAN PULSES
% =====================================================

%\foreach \x in {3.0,5.6,9.3}
%{

% -------------------------------
% Rising edge
% -------------------------------

%\draw[
%olive,
%thick,
%smooth,
%samples=100,
%domain=-0.90:-0.40,
%variable=\u
%]
%plot
%(
%{\x+\u},
%{2*exp(-18*(\u+0.40)^2)}
%);

% -------------------------------
% Flat top
% -------------------------------

%\draw[olive,thick]
%({\x-0.40},2)
%--
%({\x+0.40},2);

% -------------------------------
% Falling edge
% -------------------------------

%\draw[
%olive,
%thick,
%smooth,
%samples=100,
%domain=0.40:0.90,
%variable=\u
%]
%plot
%(
%{\x+\u},
%{2*exp(-18*(\u-0.40)^2)}
%);

%}

% =====================================================
% FILLED SNAP PULSE
% =====================================================

\foreach \x in {3.7,6.8,11.45}
{

\fill[olive!70,fill opacity=0.30]

({\x-0.90},0)

-- plot[
smooth,
samples=100,
domain=-1.1:-0.6,
variable=\u
]
(
{\x+\u},
{2*exp(-18*((\u+0.60)*(\u+0.60)))}
)

-- ({\x+0.60},2)

-- plot[
smooth,
samples=100,
domain=0.60:1.1,
variable=\u
]
(
{\x+\u},
{2*exp(-18*((\u-0.60)*(\u-0.60)))}
)

-- ({\x+0.90},0)

-- cycle;

% =====================================================
% OUTLINE
% =====================================================

% Left Gaussian edge
\draw[
olive,
thick,
smooth,
samples=100,
domain=-1.1:-0.6,
variable=\u
]
plot
(
{\x+\u},
{2*exp(-18*((\u+0.60)*(\u+0.60)))}
);

% Flat top
\draw[olive,thick]
({\x-0.60},2)
--
({\x+0.60},2);

% Right Gaussian edge
\draw[
olive,
thick,
smooth,
samples=100,
domain=0.6:1.1,
variable=\u
]
plot
(
{\x+\u},
{2*exp(-18*((\u-0.60)*(\u-0.60)))}
);

}

\node at (8.35,1.0) { $\ldots$};

\end{scope}

% =====================================================
% LEGEND
% =====================================================

% ----- Displacement legend -----

\draw[
purple,
thick
]
(-3.0,-5.0)
--
(-4.0,-5.0);

\node[anchor=west]
at (-2.8,-5.0)
{\small Displacement $D(\alpha)$};

\node[anchor=west]
at (-2.8,-5.5)
{\small (Gaussian pulse)};

% ----- SNAP legend -----

\draw[
olive,
thick
]
(1.0,-5.0)
--
(2.0,-5.0);

\node[anchor=west]
at (2.2,-5.0)
{\small SNAP gate $S(\Vec{\theta})$};

\node[anchor=west]
at (2.2,-5.5)
{\small (flat-top Gaussian pulse)};

\end{tikzpicture}
}

\caption{\RaggedRight Universal control of a cavity bosonic mode using displacement and SNAP gates. An arbitrary unitary operation in a $d$-dimensional cavity subspace can be decomposed into a sequence of $k+1$ displacement gates, $D(\alpha)$,  and $k$ SNAP gates, $S(\vec{\theta})$. In this work, displacement operations are implemented using short Gaussian-shaped control pulses applied through cavity drive, while SNAP gates are realized using longer flat-top Gaussian pulses that induce photon-number-dependent rotations of the ancilla qubit. Together, these operations provide universal control over the cavity Hilbert space.} \label{fig:1}
\label{sequence}

\end{figure*}
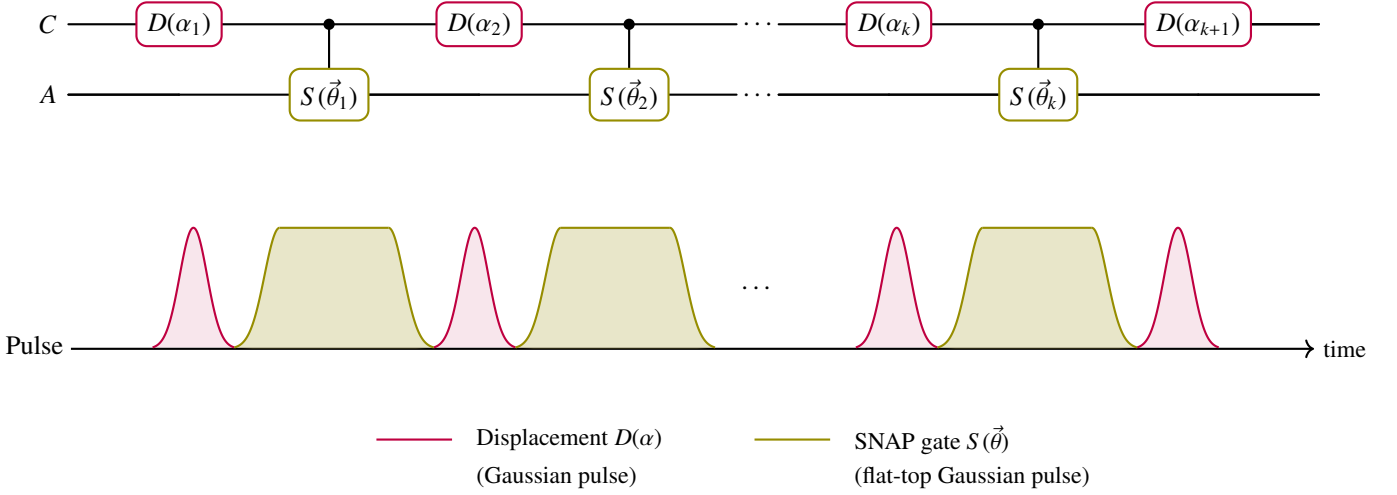

\section{Bosonic encoding and universal quantum control}
%%\label{}
The circuit QED setup consists of a transmon qubit coupled to a single mode of a microwave cavity. If the detuning between the qubit and cavity frequencies is much larger than their coupling strength, the system can be described by the dispersive Hamiltonian \cite{Blais:2020wjs}
\begin{equation}
    H_0 = \omega_c a^\dagger a + \omega_q \ket{e}\bra{e} - \chi a^\dagger a \ket{e} \bra{e},
\end{equation}
where $\omega_c$ and $\omega_q$ denote the cavity and qubit frequencies, respectively, $a(a^\dagger)$ is the cavity annihilation (creation) operator and $\ket{e}$ is the qubit excited state. The third term in the Hamiltonian represents the coupling between the cavity and the qubit, characterized by the dispersive shift $\chi$. In the dispersive regime, the transmon can be utilized as an ancillary element for controlling the cavity mode through a sequence of SNAP and displacement gates.  In this scheme, independent drives are applied to both the ancilla and the cavity. The driving Hamiltonian in the laboratory frame can be written as 
\begin{align}
    H_D =& H_1 + H_2,
\end{align}
where $H_1$ and $H_2$ represent cavity and ancilla drives respectively:
\begin{align}
    H_1 =& \epsilon_1(t)(e^{-i(\omega_1 t + \phi_1)}a^\dagger+e^{i(\omega_1 t + \phi_1)}a), \\ \nonumber
    H_2 =& \epsilon_2(t)(e^{-i(\omega_2 t + \phi_2)}\sigma^+ +e^{i(\omega_2 t + \phi_2)}\sigma^{-}).
\end{align} 
Here, $\epsilon_{1(2)}$ is the shaped pulse amplitudes, $\omega_{1(2)}$ is the drive frequency and $\phi_{1(2)}$ is the phase for the cavity (ancilla) drive. 

We now move to a frame rotating with cavity and ancilla frequencies using the unitary $U= e^{-i \omega_c a^\dagger a t} e^{-i \omega_q \ket{e}\bra{e}t}$. In this rotating frame, the total Hamiltonian $H_0 + H_D$ becomes
\begin{align}
    H_R=& U^\dagger(H_0+ H_D)U+ i\dot{U}^\dagger U  \nonumber \\
    =& - \chi a^\dagger a \ket{e} \bra{e} + \epsilon_1(t)(e^{-i(\delta_1 t + \phi_1)}a^\dagger+e^{i(\delta_1 t + \phi_1)}a) \nonumber \\
    & + \epsilon_2(t)(e^{-i(\delta_2 t + \phi_2)}\sigma^+ +e^{i(\delta_2 t + \phi_2)}\sigma^{-}), 
\end{align}
where $\delta_1 = \omega_c- \omega_1$ and $\delta_2 = \omega_q- \omega_2$ are the detunings of the drive tones from cavity and ancilla frequencies respectively. 

With the cavity drive, we can implement displacement operations described by the operator
\begin{equation}
    D(\alpha) = e^{\alpha a^\dagger - \alpha a}, 
\end{equation}
where $\alpha = - i\int dt \epsilon_1(t) e^{-i \phi_1}$. The displacement operation alone is not sufficient to generate arbitrary cavity states and therefore does not provide universal control \cite{Ma:2021rum}. The dispersively coupled ancilla introduces the required nonlinearity into the system, enabling indirect control of the cavity mode through the SNAP gate defined as
\begin{equation} \label{snap-gen}
    S(\Vec{\theta}) = \sum_{n= 0}^{ d-1} e^{i \theta_n} \ket{n} \bra{n},
\end{equation}
which imparts arbitrary phases $\Vec{\theta} = \{\theta_n\}_{n= 0}^ {d-1}$ to the Fock states within the truncated $d$-dimensional cavity subspace. 

To implement SNAP gate, the ancilla is driven weakly ($|\epsilon_2| \ll \chi$), such that the accumulated geometric phase imparts a phase $\theta_n$ conditioned on the cavity Fock state $\ket{n}$. In particular, to apply a phase factor $e^{i \theta_n}$ to the Fock state $\ket{n}$, a sequence of two $\pi$ pulses is applied to the ancilla. The first $\pi$ pulse is applied with detuning $\delta_2 = n \chi$ and phase $\phi_2= 0$, thereby driving the transition $\ket{n, g} \rightarrow \ket{n, e}$ between the ancilla states. A second $\pi$ pulse, with same detuning but phase $\phi_2= \pi - \theta_n$ drives the state back to $\ket{n, g}$ but about an axis rotated by an angle $\pi - \theta_n$ relative to the first pulse. As a result, the transformation $\ket{n, g} \rightarrow e^{i \theta_n} \ket{n, g}$ is realized. The acquired phase $\theta_n$ is geometric in nature and is proportional to the solid angle subtended on the Bloch sphere by the evolution trajectory of the effective two-level subsystem $\{\ket{n,g}, \ket{n, e}\}$.    
If the cavity is initially prepared in a superposition of Fock states $\ket{\psi}_c = \sum_{k=0}^{d-1} c_k \ket{k}$, then the above sequence of operations results in
\begin{equation}
    \ket{\psi}_c \ket{g} = \sum_{k=0}^{d-1} c_k \ket{k} \ket{g}\rightarrow \big(e^{i \theta_n} c_n \ket{n} + \sum_{k \neq n} c_{k} \ket{k}\big) \ket{g},
\end{equation}
thereby implementing a single-angle SNAP gate on the Fock state $\ket{n}$ 
\begin{equation} \label{single-snap}
    S_n(\theta_n) = e^{i \theta_n} \ket{n} \bra{n}.
\end{equation}
More generally, a multi-angle SNAP gate of the form given in Eq.\eqref{snap-gen}: $S(\Vec{\theta}) = \prod_{i= 0}^{d-1} S_i(\theta_i)$, can be realized by applying a pulse containing multiple frequency components, thereby generating conditional rotations in different ancilla subspaces simultaneously \cite{Bornman:2024qdm}. 

It can be shown that universal control over cavity mode can be achieved by combining displacement operations with SNAP gates \cite{Krastanov:2015rkt}. For instance, consider the unitary operation that performs a rotation by an angle $\xi$ within the Fock subspace $\{\ket{n}, \ket{n+1}\}$ :
\begin{align}
    U_{n,n+1}(\xi) = & \cos \xi (\ket{n} \bra{n} + \ket{n+1} \bra{n+1}) + \sin \xi (\ket{n} \bra{n+1} \nonumber \\ & - \ket{n+1} \bra{n}).
\end{align}
Using a sequence of displacement and SNAP gates $U_{n, n+1}(\xi)$ can be implemented as
\begin{equation} \label{U01}
    U_{n,n+1} (\xi) \approx D(\alpha) S_n(\Vec{\theta}_n) D(-2 \alpha) S_n(\Vec{\theta}_n) D(\alpha),
\end{equation}
where $\Vec{\theta}_n = (\underbrace{\pi, \dots, \pi}_{n}, 0, \dots)$ denotes the SNAP-angle vector. Thus $S_n(\Vec{\theta}_n)$ imparts a phase of $\pi$ to the lowest $n$ Fock states, $\{\ket{k}\}_{k= 0}^{n-1}$. The displacement parameter $\alpha$ can, in general, be determined numerically through optimization. However, Ref.~\cite{Job:2023huu} demonstrated an analytical approach for obtaining $\alpha$ as a function of the rotation angle $\xi$. For small $\xi$, both sides of Eq.~\eqref{U01} can be expanded perturbatively to first order in $\xi$ and $\alpha$, yielding
\begin{equation} \label{alpha-approx}
    \alpha = -\frac{\xi}{4 \sqrt{n+1}}.
\end{equation} 
The corresponding gate infidelity scales as $O(\xi^6)$, indicating that this approximation remains highly accurate for small rotation angles.

In general, an arbitrary unitary operation within the $d$-dimensional cavity subspace can be implemented by interleaving $k+1$ displacement operations with $k$ SNAP gates  (Fig. \ref{fig:1}):
\begin{equation} \label{unit-sequence}
    U_d =  D(\alpha_{k+1}) S(\Vec{\theta}_k)D(\alpha_k)...S(\Vec{\theta}_1)D(\alpha_1).
\end{equation}
Each displacement-SNAP layer contributes $d+1$ real parameters to the decomposition. Since an arbitrary $d$-dimensional unitary is characterized by $d^2$ real parameters, the number of required displacement-SNAP layers scales approximately as $k \sim O(d)$. Nevertheless, by employing advanced optimization techniques \cite{Fosel:2020oyj, Kudra:2021gbu}, high-fidelity cavity control can be achieved even with relatively short sequences consisting of only $3\text{--}4$ displacement-SNAP layers.

\section{Simulating two-flavor oscillations}
%%\label{}
Quantum simulation of two-flavor neutrino oscillations involves a two-dimensional Hilbert space, which can be encoded using the two lowest Fock states of the cavity, $\{ \ket{0}, \ket{1}\}$. In the two-flavor vacuum oscillation framework, neutrino mixing is described by a unitary matrix parameterized by the mixing angle $\theta_{vac}$:
\begin{equation} \label{mixing-matrix}
    \mathscr{U}(\theta_{vac}) = \begin{pmatrix}
        \cos \theta_{vac} & \sin \theta_{vac} \\
        -\sin \theta_{vac} & \cos \theta_{vac}
    \end{pmatrix}.
\end{equation}
If the neutrino is initially prepared in the flavor state $\ket{\nu_e}$, it undergoes flavor oscillations during propagation. The corresponding survival probability amplitude is given by \cite{Giunti:2007ry} 
\begin{equation} \label{amplitude}
    A(\nu_e \rightarrow \nu_e) = \bra{\nu_e} \mathscr{U}(\theta_{vac}) \mathscr{U}(\phi) \mathscr{U}^\dagger(\theta_{vac}) \ket{\nu_e},
\end{equation}
where $\mathscr{U}(\phi)$ describes the propagation-induced phase evolution in the mass basis and is given by
\begin{equation} \label{phase-matrix}
    \mathscr{U}(\phi) = \begin{pmatrix}
        e^{- i \phi/2} & 0 \\
        0 & e^{i \phi/2}
    \end{pmatrix}.
    \end{equation}
Here, $\phi = \Delta m^2 L/2E$, where $\Delta m^2= m_2^2- m_1^2$ is mass-squared difference, $L$ is the propagation distance and $E$ is the neutrino energy. The survival probability is then obtained from
\begin{equation}\label{2-flav-prob}
    P(\nu_e \rightarrow \nu_e) = |A(\nu_e \rightarrow \nu_e)|^2.
\end{equation}

\begin{figure}
\centering
   \includegraphics[width=1\linewidth]{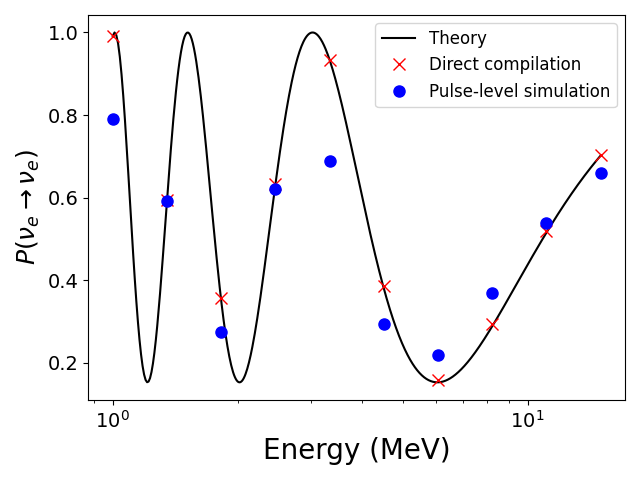}
   \caption{\RaggedRight Quantum simulation of two-flavor neutrino vacuum oscillation for the Fock-state-encoded cavity qubits. The solid line represents the theoretical prediction ($\theta_{vac} = 33.5^\circ, \Delta m^2= 7.5 \times 10^{-5}~ \mbox{eV}^2$), while the symbols correspond to the quantum simulation results obtained using direct compilation [Eq.~\eqref{2-flav-gate}] and pulse-level simulation [\ref{apprex-1}].}\label{fig:earth-prob} \label{fig:2}
\end{figure}

Using the Fock basis encoding of the cavity modes, the matrix $\mathscr{U}(\theta_{vac})$ can be compiled using the displacement-SNAP sequence described in Eq.~\eqref{U01}
\begin{equation}
    \mathscr{U}(\theta_{vac}) \approx D(\alpha_{vac}) S_0(\pi) D(-2 \alpha_{vac}) S_0(\pi) D(\alpha_{vac}).
\end{equation}
Here, $S_0(\pi)$ denotes a single-angle SNAP gate \eqref{single-snap} that imparts a phase shift of $\pi$ to the Fock state $\ket{0}$:
\begin{equation} \label{snap-pi}
    S_0 (\pi) = - \ket{0} \bra{0} + \sum_{k=1}^{d-1} \ket{k} \bra{k}.
\end{equation}
For the displacement parameter $\alpha_{vac}$, we employ the approximation given in Eq.~\eqref{alpha-approx}, 
\begin{equation}
    \alpha_{vac} = -\frac{\theta_{vac}}{4}.
\end{equation}
Direct compilation of $\mathscr{U}(\theta_{vac})$ using this value of $\alpha$ yields an infidelity below $10^{-4}$. The phase evolution matrix in Eq.~\eqref{phase-matrix} can be implemented using a generic SNAP gate \eqref{snap-gen} with the angle vector 
\begin{equation}
    \Vec{\theta}_\phi = (-\phi/2, \phi/2, 0, \dots)
\end{equation}
Thus, the overall unitary evolution of the neutrino flavor states in Eq.~\eqref{amplitude} can be approximated using the following displacement-SNAP sequence:
\begin{align} \label{2-flav-gate}
    U_{d}^{{cav}} = & D(\alpha_{vac}) S_0(\pi) D(-2 \alpha_{vac}) S_0(\pi) D(\alpha_{vac}) \nonumber \\
   &  S(\Vec{\theta}_\phi)D(-\alpha_{vac}) S_0(\pi) D(2 \alpha_{vac}) S_0(\pi) D(-\alpha_{vac}).
\end{align}
The unitary $U_{d}^{{cav}}$ denotes the effective operation implemented in the truncated $d$-dimensional Fock subspace of the cavity. Measurement of the cavity state then yields the probability in Eq.~\eqref{2-flav-prob}: $|\bra{0}U_{d}^{{cav}}\ket{0}|^2$. In Fig.~\ref{fig:2}, we plot the resulting survival probability for neutrinos initially prepared in the flavor state $\ket{\nu_e}$. In the direct-compilation approach, the matrix representations of the displacement and SNAP operations are used with the prescribed gate parameters. In the pulse-level simulation, the displacement and SNAP gates are implemented using shaped control pulses, as discussed in ~\ref{apprex-1}. The overall gate fidelity differs significantly in the two cases. In the direct-compilation case, which neglects dissipation, we obtain a gate fidelity between $0.990$ and $0.995$. In contrast, the pulse-level simulations yield gate fidelities of approximately $0.97$ for most data points. Despite this reduction in fidelity, both approaches produce oscillation probabilities that are in close agreement with the corresponding theoretical predictions.   

\section{Simulating three-flavor oscillations}
Three-flavor neutrino oscillations are described by a $3\times3$ mixing matrix parametrized by the three angles $\theta_{12}, \theta_{23}$ and $\theta_{13}$
\begin{equation}\label{3-flav}
    \mathscr{U}_{3f}= R_{23}S(\delta)R_{13}S^\dagger(\delta)R_{12}.
\end{equation}
Here, the matrices $R_{ij}$ represent rotations by an angle $\theta_{ij}$ in the $(i,j)$ subspace, while $S(\delta) = \mbox{Diag} (e^{-i \delta/2}, 1, e^{i \delta/2})$,
where $\delta$ denotes the Dirac CP-violating phase.  The neutrino flavor oscillation probability is given by
\begin{equation} \label{3-flav-prob} 
    P(\nu_\alpha \rightarrow \nu_\beta) = |\bra{\nu_\beta}\mathscr{U}_{3f} \mathcal{U}\mathscr{U}_{3f}^\dagger\ket{\nu_\alpha}|^2,
\end{equation}
The matrix $\mathcal{U}$ is the time-evolution matrix given by 
\begin{equation}
    \mathcal{U} = \mbox{Diag}(1, e^{i \phi_1}, e^{i \phi_2}),
\end{equation}
where $\phi_1 = \Delta m^2_{21}L/2E$ and $\phi_2= \Delta m^2_{32}L/2E $, with $\Delta m^2_{21}= m_2^2- m_1^2$ and $\Delta m^2_{32} = m_3^2 - m_2^2$ representing the mass-squared differences.

Direct compilation of Eq.~\eqref{3-flav} would require approximately $12$ SNAP gates. Consequently, evaluating the oscillation probability in Eq.~\eqref{3-flav-prob} would require nearly $25$ SNAP gates in total. Since each SNAP gate has a typical duration of $\sim 5~\mu\text{s}$, the complete pulse sequence would require approximately $125~\mu\text{s}$. While this duration remains well below the cavity lifetime $\sim 1$ ms, it is comparable to the typical coherence time of the ancilla ($\sim 100~ \mu$s). As a result, direct pulse-level compilation is not a practical approach for implementing the full three-flavor oscillation protocol.

To overcome this limitation, we adopt the optimized compilation scheme proposed in \cite{Fosel:2020oyj}, which substantially reduces the number of required displacement-SNAP operations. We implement a variation of the optimization procedure described in Ref.~\cite{Fosel:2020oyj} using the \texttt{qgrad} package\footnote{\url{https://github.com/qgrad/qgrad}}, as discussed in ~\ref{apprex-2}. This approach enables the construction of short gate sequences that accurately approximate the target unitary while remaining compatible with the coherence constraints of current circuit QED devices. 

\begin{figure}
\centering
   \includegraphics[width=1\linewidth]{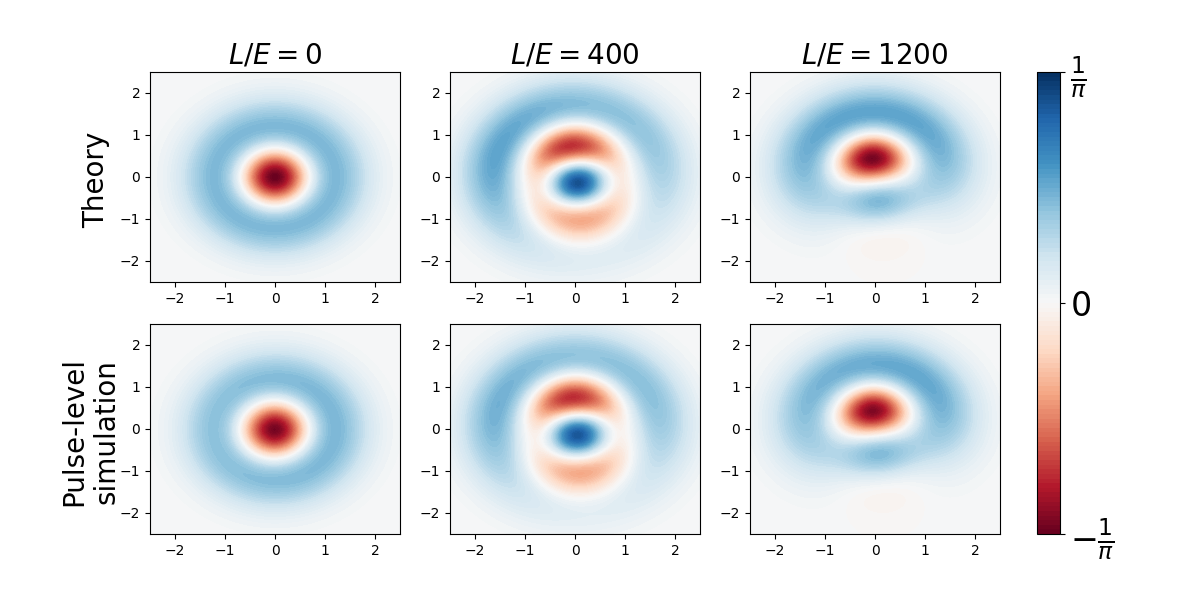}
   \caption{\RaggedRight Wigner function of the neutrino flavor state $\ket{\nu_\mu}$, initially encoded in the cavity Fock state $\ket{1}$. As the neutrino propagates, flavor mixing causes the encoded cavity state to evolve. The top row shows the theoretically predicted states, while the bottom row shows the corresponding pulse-level simulation results.}   \label{fig:3}
\end{figure}

\begin{figure}
\centering
   \includegraphics[width=1\linewidth]{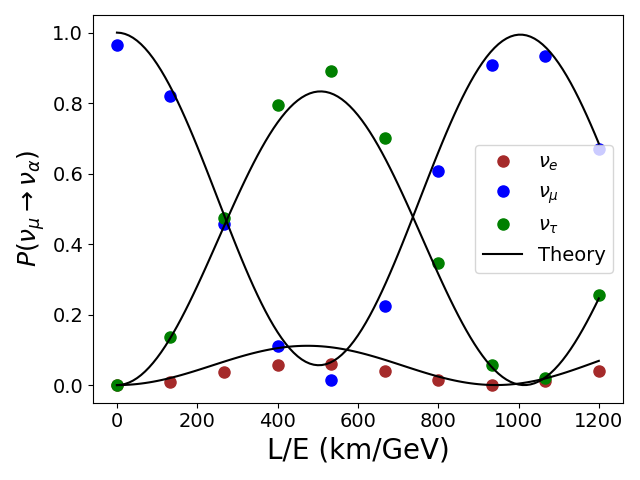}
   \caption{\RaggedRight Pulse-level simulation of three-flavor neutrino oscillations. The three flavor states are encoded in the cavity Fock subspace $\{ \ket{0}, \ket{1}, \ket{2}\}$. The neutrino mixing matrix is compiled using two building blocks of the form \eqref{build-block} and the parameters of the SNAP and displacement gates are optimized numerically using the \texttt{qgrad} package.}   \label{fig:4}
\end{figure}

In Figures \ref{fig:3} and \ref{fig:4}, we present the results of the pulse-level simulation of three-flavor neutrino oscillations. The neutrino is initially prepared in the flavor state $\ket{\nu_\mu}$, which is encoded in the cavity Fock state $\ket{1}$. As the neutrino propagates, flavor mixing causes the state $\ket{\nu_\mu}$ to evolve into a superposition of the three flavor states. Within the cavity encoding, this corresponds to the evolution of the initial Fock state into a superposition of the basis states $\ket{0}$, $\ket{1}$, and $\ket{2}$. This evolution is visualized in Fig.~\ref{fig:3} through the Wigner functions of the cavity mode at different values of $L/E$. At $L/E=0$, the Wigner function corresponds to that of the Fock state $\ket{1}$, reflecting the initial $\nu_\mu$ state. As $L/E$ increases, the Wigner distribution gradually deforms, indicating the development of coherent superpositions among the cavity Fock states. The strongest mixing occurs near $L/E = 400$ km/GeV, where the cavity state contains significant contributions from all three basis states $\ket{0}$, $\ket{1}$, and $\ket{2}$. This is manifested by the appearance of interference features in the Wigner distribution that are absent in the initial state.

The corresponding oscillation probabilities are shown in Fig.~\ref{fig:4}. The pulse-level simulation results are found to be in close agreement with the theoretical predictions over the entire range of $L/E$ values considered. Using direct compilation, we obtain gate fidelities close to $0.99$, whereas in the pulse-level simulations the fidelities decrease to approximately $0.925$. Despite this reduction in fidelity, the overall agreement demonstrates that the optimized displacement-SNAP sequences accurately reproduce the target neutrino oscillation dynamics.

\section{Summary and conclusions}
%%\label{}
With recent advances in quantum hardware, a wide variety of quantum encoding and control schemes are being actively explored. In this work, we presented quantum simulation of two- and three-flavor neutrino oscillations using bosonic encoding of a cavity mode. The neutrino flavor states were mapped onto the Fock states of a microwave cavity and controlled using sequences of displacement and SNAP gates. By employing both analytical and numerical optimization techniques, we obtained high-fidelity implementations of the required unitary operations. The resulting oscillation probabilities, obtained from direct compilation and pulse-level simulations, show good agreement with the theoretical predictions. 

These results highlight the potential of cavity-based bosonic encodings as a promising platform for the quantum simulation of particle-physics phenomena. The large Hilbert space of a single cavity mode enables efficient encoding of quantum information into qudits. Moreover, the ability to implement complex unitary operations using only two control channels provides a scalable and hardware-efficient approach to quantum control. These features make cavity-based bosonic encoding particularly attractive for the quantum simulation of many-body systems, lattice gauge theories, and other challenging problems in high-energy physics.

%% The Appendices part is started with the command \appendix;
%% appendix sections are then done as normal sections
\appendix 

\section{Pulse-level simulation}\label{apprex-1}
To simulate the circuit QED dynamics and evaluate the neutrino oscillation probabilities, we employ the master equation
\begin{equation}
    \frac{\partial \rho}{\partial t} = - i[H_R, \rho] + \frac{1}{T_1} \mathcal{D}(\sigma_-) \rho +  \frac{1}{T_\phi} \mathcal{D}(\sigma_z) \rho,
\end{equation}
where $T_1$ and $T_\phi$ denote the relaxation and pure dephasing times of the ancilla, respectively, while, $\mathcal{D}$ represents the Lindblad superoperator defined as $\mathcal{D}(a) \rho = a \rho a^\dagger - \frac{1}{2} \{a^\dagger a, \rho\} $ for a given operator $a$. Since current-generation three-dimensional microwave cavities exhibit high coherence times on the order of $\sim 1$ ms \cite{krasnok2024superconducting}, decoherence effects associated with the cavity can be safely neglected in our simulations. The master-equation simulations were performed using \texttt{QuTiP} \cite{qutip5}.  

At the pulse level, the compilation of an arbitrary unitary operation is achieved through the sequential application of displacement and SNAP operations, as described in Eq.~\eqref{unit-sequence}, using shaped control waveforms. Consequently, the control protocol requires that the cavity and ancilla are never driven simultaneously. In our simulations, the cavity Hilbert space is truncated at a cutoff dimension $d=6$. The displacement operation is implemented using a Gaussian pulse of duration $t_{c} = 10$ ns and standard deviation $\sigma = t_{c}/5$. The pulse amplitude is chosen to be purely imaginary in order to realize the desired displacement parameter $\alpha = - i \int_{t_1}^{t_2} \epsilon_1(t) dt$, where $t_1$ and $t_2$ denote the initial and final times of the pulse interval, respectively. Furthermore, during the cavity drive, the ancilla must remain in its ground state $\ket{g}$. 

The implementation of SNAP gate is carried out using a flat-top Gaussian pulse with pulse amplitude $|\epsilon_2| = \chi/10$, rise time $t_r = 1000$ ns and $\sigma =t_r/3$. This operation consists of a sequence of two consecutive $\pi$-pulses such that $\int_{t_1}^{t_2} \epsilon_2(t)dt = \pi$. The total duration of each $\pi$ pulse is $t_a = t_2- t_1  \approx 2.77 \mu$s. We first consider the single-angle SNAP gate \eqref{snap-pi} which imparts a phase shift of $\pi$ to the Fock state $\ket{0}$. The first $\pi$-pulse with detuning $\delta_2= 0$ and phase $\phi_2= 0$ drives the ancilla transition $\ket{g} \rightarrow \ket{e}$, corresponding to a rotation of the ancilla state about the $X$-axis of the Bloch sphere. The second identical $\pi$-pulse induces another $X$-rotation, driving the transition $\ket{e} \rightarrow \ket{g}$. The resulting closed trajectory encloses a solid angle of $2\pi$ giving rise to a geometric phase factor $e^{i\pi}$. Consequently, the transformation $\ket{0, g} \rightarrow e^{i \pi} \ket{0, g}$ is realized, thereby implementing the SNAP gate in Eq.~\eqref{snap-pi}.

The multi-angle SNAP gate in Eq.~\ref{snap-gen} can be realized through multiple ancilla rotations, each conditioned on a specific cavity Fock state. These conditional rotations may be implemented either simultaneously or sequentially \cite{Bornman:2024qdm}. However, when the number of frequency components in the SNAP gate is relatively small, simultaneous driving is generally preferable, as it reduces the overall gate duration.

Ideally, the total pulse duration should remain well within the coherence time of the ancilla qubit. For the implementation of two-flavor neutrino oscillations, the protocol requires a total of three SNAP gates and four displacement operations, as shown in Eq. \eqref{2-flav-gate}. This corresponds to an overall gate duration of approximately $t_g \approx 21~ \mu$s. State-of-the-art transmon qubits have demonstrated coherence times exceeding 500 $\mu$s \cite{Bal:2023ccn, Wang:2021vvz}, although more typical experimental values lie in the range $T_1\sim100-200~ \mu$s. In our simulations, we consider $T_1 = 100~ \mu$s and $T_2= 50~ \mu$s, corresponding to a pure dephasing time of $T_\phi \approx 55~ \mu$s. Since the total gate time $t_g$ remains significantly smaller than both $T_1$ and $T_\phi$, the neutrino oscillation dynamics can be simulated with reasonable fidelity.
%% \label{}

\section{Numerical optimization of gate parameters}\label{apprex-2}
The compilation procedure proposed in Ref. \cite{Fosel:2020oyj} is based on building blocks of the form
\begin{equation}\label{build-block}
    B(\alpha, \Vec{\theta}) = D^\dagger(\alpha) S(\Vec{\theta}) D(\alpha). 
    \end{equation} 
These building blocks are assembled hierarchically to approximate a desired target operation. The implementation proceeds in two stages: an initialization stage followed by a fine-tuning stage. During the initialization stage, the building blocks are inserted hierarchically to construct a coarse approximation of the target operation. The objective of this stage is to iteratively improve the overlap with the target unitary while determining the required sequence length $T$ of the building-blocks. In the subsequent fine-tuning stage, the gate parameters are optimized using gradient-descent techniques to obtain a high-fidelity gate operation.

In our implementation, we use a simplified version of the method described above. We first compile the unitary in Eq.~\eqref{3-flav} using a fixed number of building blocks, considering the cases $T=2$ and $T=3$. The parameters of the displacement and SNAP gates are then optimized using the \texttt{qgrad} package. The optimization minimizes a cost function defined by the infidelity between the target unitary and the compiled unitary, with gradients computed via JAX-based automatic differentiation. For the case $T=2$, we obtain following parameter values after optimization
\begin{align*}
    & \alpha_1 = -0.14083214, \alpha_2= 0.1396543 \\
    & \Vec{\theta}_1 = ( 1.2601506, -1.8827367,  1.2575893,   1.2569728,\\& \hspace{1.7cm}  1.2566546,
         1.2565293) \\
    & \Vec{\theta}_2 = ( -1.0190855,  2.1238182, -1.0165251,  -1.0158992, \\& \hspace{1.7cm} -1.0155833,-1.0154592)
\end{align*}
For both $T=2$ and $T=3$, the optimization yields gate fidelities exceeding 0.99 at the unitary-compilation level. However, in pulse-level simulations the fidelities decrease to approximately 0.925 and 0.885, respectively, owing to the accumulation of errors associated with the larger number of control pulses.

%% \label{}

%% If you have bibdatabase file and want bibtex to generate the
%% bibitems, please use
%%
\bibliographystyle{elsarticle-num}
\bibliography{zznuosc, zzquantum}

%% else use the following coding to input the bibitems directly in the
%% TeX file.

%%\begin{thebibliography}{00}

%% \bibitem[Author(year)]{label}
%% For example:

%% \bibitem[Aladro et al.(2015)]{Aladro15} Aladro, R., Martín, S., Riquelme, D., et al. 2015, \aas, 579, A101

%%\end{thebibliography}

\end{document}